\documentclass[12pt]{article}
\usepackage{sproclmo}
\usepackage{epsfig}
 \def \(({\left(}
\def \)){\right)}
\begin{document}

\bibliographystyle{unsrt} 
\title{AN ATOMIC LINEAR STARK SHIFT VIOLATING P BUT NOT T ARISING FROM THE ELECTROWEAK\\ NUCLEAR
ANAPOLE MOMENT  }

\author{M. A. BOUCHIAT   }

\address{Laboratoire Kastler Brossel \footnote {Laboratoire de l'Universit\'e Pierre et Marie
Curie et de l'Ecole Normale Sup\'erieure, associ\'e au CNRS (UMR 8552)}, \\ 24, rue
Lhomond, F-75231 Paris Cedex 05, France  } 

\author{C. BOUCHIAT }
\address{Laboratoire de Physique Th\'eorique de l'Ecole Normale Sup\'erieure
\footnote {UMR 8549: Unit\'e Mixte du Centre National de la Recherche Scientifique
 et de l'\'Ecole Normale Sup\'erieure}\\
24, rue Lhomond, F-75231 Paris Cedex 05, France }

\maketitle\abstracts{ 
We propose a direct method of detection of the nuclear anapole moment. It is 
based on the existence of a linear Stark shift for alkali atoms in
their ground state perturbed by a quadrupolar interaction of uniaxial symmetry
around a direction $\hat n$ and a magnetic field. This shift is characterized by the T-even
pseudoscalar $(\hat n
\cdot \vec B)(\hat n \wedge \vec E \cdot \vec B)/ B^2 $. It involves on the one hand the anisotropy of the
hyperfine interaction induced by the quadrupolar interaction and, on the other, the static electric dipole
moment arising from electroweak interactions inside the nucleus. The case of ground state Cs
atoms trapped in a uniaxial (hcp) phase of solid $^4$He is examined. From an explicit
evaluation of both the hyperfine structure anisotropy and the static dipole deduced from recent empirical data about
the Cs nuclear anapole moment, we predict the Stark shift. It is three times the experimental upper bound to be set on
the T-odd Stark shift of free Cs atoms in order to improve the present limit on the electron EDM.   }
\section*{Introduction}
It has been well demonstrated that parity violation in atomic transitions can be used to test
electroweak theory
\cite{bou97}. In this way, the Standard Model has been confirmed convincingly in the domain
of low energies. At present, refinements in experiments and theory allow more precise
measurements to look for a breakdown of the Standard Model predictions and hence, new
physics~\cite{ben99,mor00,fay99,ram99}. The essential parameter extracted from atomic parity violation (PV)
measurements is the weak nuclear charge
$Q_W$. This electroweak parameter appears in the definition of the dominant electron-nucleus PV
potential induced by a $Z_0$ exchange:
\begin{equation}
V_{pv}^{(1)}(r) = G_F/\sqrt{2} \cdot  Q_W/2 \cdot \gamma_5 \cdot P_V(r) \;,
\end{equation}
where the $Z_0$ couples to the nucleus as a vector particle, just as the photon does in the
Coulomb interaction. In this $Z_0$ exchange, $Q_W$ plays the same role as the electric charge in the
Coulomb interaction. $\gamma_5$ is the Dirac matrix which reduces to the electron helicity, $
\vec \sigma \cdot  \vec p/ m_e c $, in the non-relativistic limit. The distribution
$ P_V(r)$ normalized to unity represents the weak charge distribution inside the nucleus. The
physical quantity measured in atomic PV experiments is a transition electric dipole moment,
$E_1^{pv}$ between states with the same parity, like the $nS_{1/2}
\rightarrow n'S_{1/2} $ transitions. In particular the $6S_{1/2} \rightarrow 7S_{1/2}$ transition in
cesium has been the subject of several experiments, the accuracy of which has been steadily
increasing with time  \cite{bou82,bou862,gil85,noe88,woo97}.

On top of this the PV electron-nucleus interaction involves also a nuclear spin-depen\-dent contribution which
can provide valuable and original information regarding Nuclear Physics. It is generated by an
interaction of the current-current type with a vector coupling for the electron and an axial coupling
for the nucleus. The associated PV potential 
$V_{pv}^{(2)}$ is given by the following
expression: 
\begin{equation}
 V_{pv}^{(2)} = G_F/ \sqrt{2} \cdot A_W /(2 I) \cdot  \vec {\alpha} \cdot \vec I \cdot P_A(r)\;,
\end{equation}
where $\vec \alpha$ is the Dirac matrix associated with the electron velocity operator, $\vec I$ the
nuclear spin and $P_A(r) $ a nuclear spin distribution normalized to unity. The weak axial moment
of the nucleus, $A_W$, receives several contributions. The most obvious one comes from the weak
neutral vector boson $Z_0$ with axial coupling to the nucleons. However, in the standard electroweak
model the coupling constants involved nearly cancel accidentally. As first pointed out by Flambaum
{\it et al.} \cite{fla80}, a sizeable contribution to $A_W$ is induced by the contamination of the atom
by the PV interactions between the nucleons which take place {\it inside} the nucleus. The concept
relevant to describe this interaction is the nuclear anapole moment \cite{zel57}.  In fact the interaction can
be interpreted simply in terms of a chiral contribution to the nuclear spin magnetization
\cite{bou911,bou99}, as illustrated in Fig 1. In other words, one can say that the PV nuclear forces
inside any stable nucleus are responsible for the nuclear anapole moment or equivalently a nuclear
helimagnetism.  The present paper addresses the problem of {\it how to detect directly this unique
static nuclear property characteristic of parity violation in stable nuclei.}

\begin{figure}
 \centerline{\epsfxsize=80mm \epsfbox{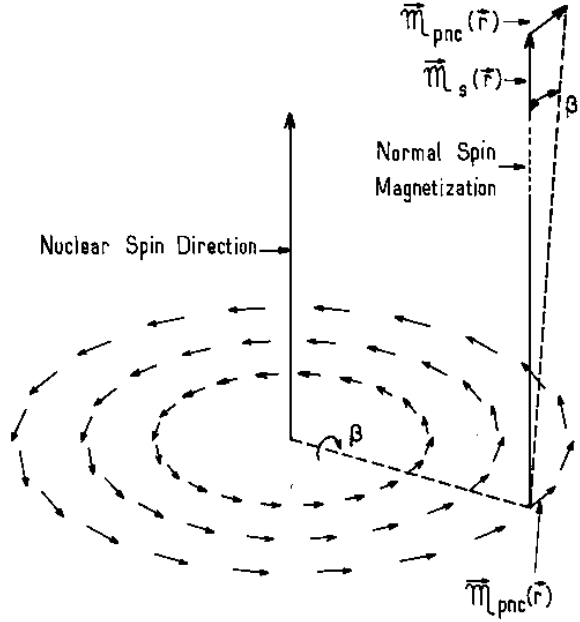}}
\caption{ \footnotesize Simplified representation of the nuclear helimagnetism (figure adapted from
C.~Bouchiat \protect \cite{bou911}). The normal spin
magnetization $\vec {\cal{ M}}_S(\vec r) $ is assumed to be a constant vector parallel to the nuclear
spin, distributed uniformely inside a sphere. Under the influence of PV nuclear forces, the   nuclear
magnetization  distribution  inside  the nucleus acquires a chiral parity non-conserving component 
$ {\vec{\cal M}} _{pnc}(\vec{r})  $, obtained by rotating $\vec {\cal{ M}}_S({\vec r})$ through the
very small angle $\beta ( r) $ around $\vec r$. Three chiral magnetization lines in the
equatorial plane are shown. The vertical normal magnetization is actually larger than $ {\vec{\cal
M}} _{pnc}(\vec{r})  $ by about six orders of 
magnitude. It can be shown \protect \cite{bou911} that the vertical anapole moment is
given, within a constant, by the magnetic moment obtained by identifying the chiral magnetization
lines with lines of electric currents.}
\end{figure}

Up to now there has been only one experimental demonstration of the nuclear anapole moment,
namely that obtained very recently by the Boulder group \cite{woo97}. In their experiment which 
gives a high precision determination of parity violation in the atomic $6S_{1/2}\rightarrow 7S_{1/2}$ Cs
transition, this effect  appears as a small relative difference, actually 
$ \sim 5\%$, between the
$E_1^{pv}$ transition dipole amplitudes measured on two different hyperfine lines belonging to
that same transition. In this case the dominant source of P without T violation comes from the electron-nucleon
$Z_0$ exchange associated with the weak charge $Q_W$ of the nucleus. This makes the extraction of the nuclear
spin-dependent part a most delicate matter. In view of the importance of this result for the determination of the
PV pion-nucleon coupling constant,
$f_{\pi}^1$ (see 
\cite{fla84}), a totally independent determination is highly desirable.

  It is well known that  T reversal invariance forbids the manifestation of
$V_{pv}^{(1)}$  in an atomic stationary state. However, we shall show in the following
sections that in such a state T reversal invariance does not forbid the manifestation of $V_{pv}^{(2)}$, hence
that of the nuclear helimagnetism. For a free atom, the rotation symmetry of the Hamiltonian leads to an exact
cancellation of the diagonal matrix elements. This property still holds true if the rotation symmetry is broken by
the application of static uniform electric and magnetic fields. However, if the symmetry is broken by the
application of a static potential of quadrupolar symmetry, for instance by trapping the atoms inside a crystal of
hexagonal symmetry, then, the stationary atomic states are endowed with a permanent electric dipole moment
which can give rise to a linear Stark shift. This offers a novel possibility of detecting the nuclear
helimagnetism having a twofold advantage:

i) In a stationary state it is {\it the sole cause of P without T violation}.

ii) It manifests itself by a modification of the atomic transition frequencies in an applied electric field,
{\it i.e. a linear Stark shift, providing for the first time an opportunity for  demonstrating the static
character of this unusual nuclear property.}

There exist in the literature other proposals for a {\it direct} detection in atoms of the nuclear
spin-dependent effect, {\it i.e.} without any participation from $V_{pv}^{(1)}$:

i) One is based on the difference between the selection rules of the potentials $V_{pv}^{(1)}$ and
$V_{pv}^{(2)}$. While the former acts as a scalar in the total angular momentum space and 
mixes only states of identical angular momentum (and opposite parity), the latter acts like a vector and
mixes states of different total angular momentum.  Consequently, one can find atomic transitions
between states of the same parity which are allowed for the nuclear spin dependent contribution but
remain forbidden for the nuclear spin independent one \cite{bou75}. One such example is the $(6p^2)
^3P_0
\to (6p^2) ^1S_0$ lead transition at 339.4 nm, strictly forbidden for even isotopes, which acquires a
non-vanishing matrix element $E_1^{pv}$ in odd isotopes owing to the PV interaction involving
the nuclear spin, which mixes the $(6p^2) ^1S_0$ state to the $(6p 7s) ^3P_1$ state of opposite parity 
\cite{bou75}.

ii) A second approach, invoked by several groups in the past and now
under serious consideration \cite{bud98}, consists in the detection of an $E_1^{pv}$ amplitude 
via a right-left asymmetry appearing in hfs transition probabilities for the ground state of 
potassium in the presence of a strong magnetic field (magnetic and hyperfine splittings of
comparable magnitude).

iii) There is also the possibility of detecting the energy difference in the NMR spectrum of
enantiomer molecules \cite{bar86}.

In view of the extreme difficulty of these other projects, we believe that, over and above its intrisic scientific 
interest, the linear Stark shift discussed in this paper deserves careful consideration.

The first section of this paper recalls the main angular momentum properties of the permanent
nuclear spin-dependent PV electric dipole operator arising from the nuclear anapole moment. In addition 
we compute its magnitude for the cesium atom using recent empirical data relative to the Cs $6S
\rightarrow 7S$ transition. The next section (sec. 2) shows that this dipole can manifest itself via a
linear Stark shift only if the free atom symmetry is broken. After this we consider the case where the atom is
perturbed by a crystal field of uniaxial symmetry. Here, the crystal axis
$\vec n$, and the applied electric and magnetic fields create a chiral environment permitting the
existence of a linear Stark shift, the explicit expression for which is given. In the section
3, we examine a realistic experimental situation where its observation looks reasonably feasible:
this deals with Cs atoms trapped inside a
$^4$He crystal matrix of hexagonal symmetry. We have investigated quantitatively how, by breaking the
atomic symmetry, the matrix
induced perturbation manages to generate a linear Stark shift. Moreover, we evaluate both the matrix induced
anisotropy and the shift. The details of the necessary calculation based on a semi-empirical method are given in
the Appendix.  In the final section we suggest another experimental approach in which the atoms are no longer
submitted to a crystal field, but are instead perturbed by an intense nonresonant radiation field.         
\section{The permanent nuclear spin-dependent PV electric dipole }
 \subsection{ Symmetry considerations }

The space-time symmetry properties of the  atomic electric dipole induced by 
the nuclear spin dependent PV interaction have
been presented before in  many  review papers (see for instance\cite{san86}).  We recall them here for
completeness, since  they constitute
the starting point of the linear Stark  shift  calculation  developed in the present paper.   

First, we wish to
 stress that the existence of the anapole moment interaction not only implies the existence of a
transition dipole proportional to the nuclear spin, but also that of an electric dipole
operator having diagonal matrix elements between stationary atomic states. 
This electric dipole is found to be  proportional to the operator $\vec s \wedge\vec I$. Therefore it does not undergo
the same transformation under P as does an ordinary dipole, since it is a pseudovector  instead of a vector. We also
note that it is even under T-reversal, so that  the quantity
$(\vec s \wedge \vec I) \cdot \vec E$, associated with a linear Stark shift, violates P, but does not violate T
invariance.   

  It is convenient to define  $ \vec{d}_{pv} (n^{\prime},n) $ as the  effective pv electric  dipole moment operator 
acting  in  the tensor product ${\cal E}_S \bigotimes {\cal E}_I$ of the electronic and nuclear angular 
momentum spaces, which describes the transition  between  two $ S_{1/2}$  subspaces corresponding to given
radial  quantum numbers $n $ and $n^{\prime}$. This effective dipole operator includes both contributions
from potentials  $ V_{pv}^{(1)}$ and $ V_{pv}^{(2)}$. Rotation
invariance  together with  the fact  that  $ V_{pv}^{(2)}$ is linear in $ \vec{I} $  implies that  $ \vec{d}_{pv}
(n^{\prime},n) $ can be written under  the following  general form: 
\begin{equation}
 \vec{d}_{pv} (n,n^{\prime}) =- i Im \, E_{1pv}^{(1)}(n,n^{\prime}) \ \vec{\sigma} +i \, a (n,n^{\prime}) \,
\vec{I}+  b (n,n^{\prime})\, \vec{s} \wedge \vec{I} \, ,
\label{dpv}
 \end{equation}
where the   real quantities  $ a(n,n^{\prime}) $ and $b(n, n^{\prime}) $
parametrize the contribution of the nuclear spin-dependent  pv potential.  Time
reversal  invariance of $ V_{pv}^{(1)}$ and $ V_{pv}^{(2)}$  implies the following  
relations under the exchange $ n \leftrightarrow n^{\prime} $:
\begin{eqnarray}
Im \, E_{1pv}^{(1)}(n,n^{\prime}) &=&-Im \, E_{1pv}^{(1)}(n^{\prime},n) \,,\nonumber \\
a(n,n^{\prime})&=&-a(n^{\prime},n) \,,\nonumber \\
 b(n,n^{\prime})&=&b(n^{\prime},n) \,.
\end{eqnarray}
The effective pv  static dipole moment $ \vec{ \cal D} _{pv}= \vec{d}_{pv} (6,6) $ relative to the
ground state is  then  given  by :
\begin{equation}
\vec{ \cal D} _{pv}= b (6,6)\, \vec{s} \wedge \vec{I}= d_I \, \vec{s} \wedge \vec{I} \, .
\label{Dpv}
\end{equation}
If we introduce the total angular momentum $\vec F=
\vec s + \vec I$, using simple relations of angular momentum algebra, one can derive the useful
identity:
\begin{equation}
\vec s \wedge\vec I \equiv  [\vec F^2 \;,\; \frac{-i}{2} \vec s] \;.
\end{equation} 
It then becomes obvious that, in low magnetic fields and without  external perturbation, 
the dipole operator $\vec {\cal D}^{pv}$ has no diagonal matrix elements between atomic eigenstates.
In fact, as demonstrated in the next section of this paper,  
a manifestation of this dipole requires special conditions for breaking the
free-atom rotational symmetry. 
\subsection{Magnitude of the permanent dipole.}
The magnitude, $d_I$, of the permanent dipole will play a decisive role
in the assessment  of the
feasibility of an experiment. We are now going to perform the evaluation   
of $d_I $ in the interesting case
of cesium. We proceed in two steps : first we compute directly   $ b(6,7)$
from experimental data,
then we  give a theoretical evaluation of the ratio $ b(6,6)/b(6,7)$.
It is convenient to use the notations of ref \cite{bou862}  and
\cite{bou911}  and to rewrite  $ \vec{d}_{pv} (6,7) $
as:
\begin{equation}
\vec{d}_{pv} (6,7)= -i \,Im E_1^{pv} ( 6,7) \(( \vec{\sigma } + \eta \frac{
\vec{I} } {I} +
i\, {\eta}^{\prime} \vec{\sigma} \wedge \frac{ \vec{I} } {I} \)) \,.
\end{equation}

The nuclear spin dependent potential  $V_{pv}^{(2)}$ induces a specific 
dependence of the pv  transition dipole on the initial
and final hyperfine quantum numbers,  $ F$ and $F^\prime$.
In order to isolate the  $V_{pv}^{(2)}$ contribution, we are led,
following ref \cite{bou862}, to the introduction of 
the reduced amplitudes $ d_{F \,F^\prime} $:
\begin{equation}
d_{F \, F^\prime}(  \eta , {\eta}^{\prime})=
\frac{\langle 7S, F^\prime \, M^\prime \vert \vec d_{pv} \vert 6S, F\, M \rangle}
{\langle F^\prime  \, M^\prime \vert \vec{\sigma}  \vert  F\, M \rangle}. 
\end{equation} 
The  amplitudes $ d_{F \,F^\prime}(\eta,{\eta}^{\prime}) $ are
 tabulated in Table XXII of ref\cite{bou862} and reduce to $-i\, Im E_1^{pv}(6,7)$
 for vanishing $\eta$ and ${\eta}^{\prime}$. 
The quantity of interest here is the 
ratio $ r_{hf}= d_{4\,3} / d_{3\,4}$ which is given, 
 to second order in $\eta$ and $ {\eta}^{\prime} $, by :
\begin{equation}
r_{hf} \simeq1-\frac{2 I+1}{I}\, {\eta}^{\prime} .
 \end{equation}
Using the empirical  value for the ratio $ r_{hf}  $ given by the last
Boulder experiment \cite{woo97}:
$ r_{hf}-1= ( 4.9 \pm 0. 7) \times 10^{-2} $, we obtain:
\begin{equation}
{\eta}^{\prime}= -\frac{7}{16} \, ( r_{hf}-1) = (-2.1 \pm 0.3)\times 10^{-2} .
 \end{equation}
 We deduce $  b(6,7) $  by  a simple identification:
\begin{equation}
   b(6,7)=Im\,E_1^{pv} ( 6,7) \frac{2}{I}\,{\eta}^{\prime} =
(1.04 \pm 0.15\times)10^{-13}  \vert e \vert a_0 \,,\label{b67}
\end{equation}
where we have used for $Im\,E_1^{pv} ( 6,7) $   the empirical value
obtained in ref \cite{woo97}:
$$
Im\,E_1^{pv} ( 6,7) =(-0.837 \pm 0.003) \times 10^{-11 } \vert e \vert a_0 .
$$
To  compute the  ratio  $ b(6,6)/b(6,7) $,  we are going to use an approximate relation, derived in ref
\cite{bou911}, which relates the   potential  $ V_{pv}^{(2)} $ to  $ V_{pv}^{(1)}$:
\begin{equation} 
V_{pv}^{(2)}(\vec{r} )= K_A \,\frac{ A_W}{ Q_W}  2 \vec{j} \cdot \frac{ \vec{I} } {I}\,
V_{pv}^{(1)}(\vec{r} ). 
\label{relVpv2Vpv1}
\end{equation}
Here $K_A $ is a constant very close to unity which depends weakly  upon the shape of the nuclear
distributions $ P_{V}(r) $ and $ P_{A}(r) $; $\vec{j} $ is the single electron angular momentum and 
since, as we shall see, only  single particle states with $ j=1/2$   are involved, we can write  
 hereafter $ 2 \vec{j} =\vec{ \sigma} $.

This relation, valid for high Z atoms  like cesium, hinges upon the fact  that
the matrix elements $ \langle n^{\prime} p_{3/2} \vert V_{pv}^{(2)} \vert n s_{1/2} \rangle $ 
involving $p_{3/2}$ states are 
much  smaller- by a factor $ 2 \times 10^{-3} $-
 than those which involve $p_{1/2}$ states, $ \langle n^{\prime} p_{1/2} \vert V_{pv}^{(2)} \vert
ns_{1/2}\rangle$ .  This is easily 
verified in the  one-particle approximation since the radial wave  functions at the surface of  the nucleus  are
very close to  Dirac  Coulomb  wave functions for an unscreened charge $ Z $. It is argued  in 
ref \cite{bou911} that this property remains  true, to the level of few $ \% $,  when  $ V_{pv}^{(2)} (\vec{r}) $ is
replaced by the non local potential $ U_{pv}^{(2)}(\vec{r}, {\vec{r}}^{\prime} )$, which describes the core polarization
effects  within the R.P.A. approximation\footnote{ To check the validity of the relation (\ref{relVpv2Vpv1}) we 
have  compared  the values for $ \eta$ and $  {\eta}^{\prime} $  obtained in this way with those deduced from 
a direct computation \cite{blu92}  of $ {\vec{d}}_{pv}(6,7)  $. The two results agree to better than $ 10 \, \% $. }.

  The contributions of  $ V_{pv}^{(i)} $ to the effective  dipole operator 
$\vec{d}_{pv} (n,n^{\prime})$ are given as the sum of the two operators:
 \begin{eqnarray}
{\vec{A} }^{(i)} &= &  P( n^{\prime} S_{1/2}) \,V_{pv}^{(i)} \, G(E_{n^{\prime}} )\, \vec{d} \,
P( n S_{1/2}) ,  \nonumber \\
{\vec{B} }^{(i)} &= & P( n^{\prime} S_{1/2}) \, \vec{d} \, G(E_{n}) \,V_{pv}^{(i)}\, 
 P( n S_{1/2}) , 
\label{AB}
\end{eqnarray}
where $ \vec{d} $ is the electric dipole operator, $ G(E_{n})=(E_{n}-H_{atom} )^{-1}$ the Green function operator relative
to the atomic hamiltonian;  $P( n S_{1/2})$  and   $P( n^{\prime} S_{1/2})$ 
 stand for the projectors upon the  subspaces   associated   
with the configurations $ n S_{1/2}$  and  $ n^{\prime} S_{1/2}$;
 $ E_{n}$ and $E_{n^{\prime}}$ are the corresponding binding  energies.
 It follows immediatly from the Wigner-Eckart theorem that 
the operators $ {\vec{A} }^{(1)} $ and $ {\vec{B} }^{(1)} $
can be written as: 
\begin{equation} 
{\vec{A} }^{(1)}=i h(n,n^{\prime} ) \,\vec{ \sigma } \; ; \;\;\; {\vec{B} }^{(1)}=i k(n,n^{\prime} ) \,\vec{ \sigma }\,. 
\label{hk}
\end{equation} 
Using  now the relation given in equation (\ref{relVpv2Vpv1}) and the commutation of $\vec{\sigma} \cdot  
\vec{I} $ with the pseudoscalar $V_{pv}^{(1)}$, one gets the following expressions for 
$ {\vec{A} }^{(2)} $ and $ {\vec{B} }^{(2)} $: 
 \begin{eqnarray}
{\vec{A} }^{(2)}&= & i\, K_A \,\frac{ A_W}{ Q_W} \, h(n,n^{\prime} )  
 \,\((\vec{\sigma} \cdot \frac{ \vec{I} } {I}\))\,\vec{\sigma} ,   \nonumber \\
{\vec{B} }^{(2)}&= & i\, K_A \,\frac{ A_W}{ Q_W} \, k(n,n^{\prime} )  
 \,\vec{\sigma} \,\((\vec{\sigma} \cdot \frac{ \vec{I} } {I}\)) .
 \end{eqnarray}
  We arrive finally at an expression for $\vec{d}_{pv} (n,n^{\prime})$ which can be used to 
compute the ratio $ b(6,6)/b(6,7) $:

\begin{eqnarray}
\vec{d}_{pv} (n,n^{\prime})&=& i ( \,\vec{\sigma}+ K_A \,\frac{ A_W}{ Q_W }\frac{ \vec{I} } {I} )\,\((
h(n,n^{\prime})+k(n,n^{\prime}) \))+ \nonumber \\
  &  &  \,\vec{\sigma}\wedge\frac{ \vec{I} } {I} \, K_A \,\frac{ A_W}{ Q_W }
  \,\(( h(n,n^{\prime})-k(n,n^{\prime}) \)) \,.
\end{eqnarray}
    Time reversal invariance  implies $ h(n,n)= -k(n,n) $  so that we can write the
 sought for  ratio $ b(6,6)/b(6,7) $ as:
\begin{equation}
\frac{b(6,6)}{b(6,7)}= \frac{  2\, h(6,6)}{h(6,7)-k(6,7)} \; .
\end{equation}
The amplitudes  $ h(6,6)\, , \, h(6,7)$   and $ k(6,7)$   can be
 computed from  the formulas  
given in  Eqs. (\ref{AB}) and (\ref{hk}).  We have used  the explicit values of the radial matrix
 elements (parity mixing and allowed electric
dipole amplitudes) for the intermediate states\footnote{We use here the fact that, as noted by several
authors, most of the sum ($\approx 98\%$) comes from the four states $6P_{1/2},
7P_{1/2},8P_{1/2},9P_{1/2}. $   } $6P_{1/2}-9P_{1/2}$ and the energy differences involved, which are
tabulated in ref. \cite{blu92} (Table IV)\footnote{Note that a misprint in table IV of ref \cite{blu92}
has caused an interchange between the contents of columns 1 and 2 of its lower half (entitled ``7S perturbed'').  }. 
We obtain in this way:
\begin{equation}
 b(6,6)/ b(6,7) = 4.152 / 1.86=2.27\,.
\end{equation}
Combining the above result with the value of $ b(6,7)$  given by equation  (\ref{b67}) we obtain the following
estimate for $ d_I$:
\begin{equation}
 d_I \simeq 2.36 \times 10^{-13} \vert e \vert a_0\; ,
\label{dI}
\end{equation}
believed to be about 15 $\%$ accurate. 

 It is of interest to compare $ d_I$ with the P-odd T-odd EDM of the Cs atom obtained from a theoretical
evaluation using the latest   experimental  upper bound for the electron EDM~\cite{com94}  : 
$$  \vert  d_e  \vert  \leq 7.5   \times \, 10^{-19} \,\vert e \vert \, a_0.  $$
Using 
 for the cesium  anti-screening factor the theoretical value~\cite{joh86}: $ 120 \pm10,$  one gets
the following upper bound for the cesium EDM, namely the experimental sensitivity to be reached for 
improving the existing bound on $ \vert  d_e  \vert$ :
\begin{equation}
 \vert d_{CsEDM} \vert \,  \leq   9.0  \times \, 10^{-17} \,\vert e \vert \, a_0\,.
\label{dCsEDM}
\end{equation}
We are going to use Eqs.(\ref{Dpv}) and (\ref{dI}) for calculating the
linear Stark shift. 
It is interesting to note here that these equations predict also the
magnitude of the pv transition dipole involved in an eventual Cs project which would be based
on the observation of hyperfine transitions in the Cs ground state, analogous to the potassium
project mentioned in the introduction (see also \cite{bud98}).  Therefore both a project of this kind and
the linear Stark shift discussed here aim at the determination of the same physical parameter, $d_I$, but only the
observation of a dc Stark shift would prove its static character. 
\section{The linear Stark shift induced by $V_{pv}^{(2)}$}
\subsection{Need for breaking the rotation symmetry of the atomic Hamiltonian}
The parity conserving spin Hamiltonian in presence of a static magnetic field $\vec B_0$ is:
\begin{equation}
H_{spin}= A\;\vec s \cdot \vec I -g_s \mu_B \vec s \cdot \vec B_0 - \gamma_I \vec I \cdot \vec
B_0 \, .
\end{equation}

From section 2, we have seen that, to
first order in the electric field, the effect of $V_{pv}^{(2)}$  in presence of an applied electric field
can be described by the following Stark Hamiltonian: 
\begin{equation}
H_{pv}^{st} = d_I\; \vec s \wedge \vec I \cdot \vec E \equiv  - d_I\;  \frac{i}{2} [\vec F^2\;,\;\vec s
\cdot \vec E] \,.
\label{HStark}
\end{equation}
We have noted that the above identity implies the vanishing of the average value of $H_{pv}^{st}$
in the low magnetic field limit. We are going to show that this null result still remains valid {\it for arbitrary
values  and orientations of the magnetic field.}
 
To do this we consider the transformation   properties of both $H_{spin}$ and $H_{pv}^{st}$ under the 
symmetry $\Theta$, defined as the product of $T $ reversal by a rotation of $\pi $ around 
the unit vector $\hat u= \vec E \wedge \vec B_0/ {\vert E B_0\vert} $, the rotation $ R(\hat u, \pi)$.
It should be stressed that the rotation $ R(\hat u, \pi)$ and the 
symmetry $\Theta$ considered here are quantum mechanical transformations   
 acting only on the spin states. The external fields are considered as real $ c-$numbers   and 
 are not affected.  One sees immediately that $H_{spin}$ is 
invariant under the  symmetry $\Theta= T\,R(\hat u, \pi) $,  while 
$H_{pv}^{st}$ changes sign. We conclude that, in order to suppress the linear Stark shift cancellation
 we have to
break the $\Theta$ symmetry. 

 This symmetry breaking can be achieved, for instance, by perturbing the atomic $S_{1/2}$ state with a
crystal field compatible with uniaxial symmetry along the unit vector $\vec n $. 
A practical realization looks feasible, since it has been demonstrated that
Cs atoms can be trapped in a solid matrix of helium having an  hexagonal symmetry \cite{kan98}
(see also section 3). In this case the alkali S state is perturbed by the Hamiltonian\footnote{ It has
been  shown \cite{kan98} that in the bubble  enclosing the cesium atom there is a  small overlap  between the
cesium and the helium orbitals. As a consequence, the axially symmetric crystal potential inside  the bubble can be
well approximated by a regular solution of the Laplace equation.}: 
\begin{equation}  
H_b( \vec{n} )={\lambda}_b \cdot (\frac{e^2}{ 2 a_0} ) \((  (\vec{\rho}\cdot \vec{n})^2-\frac{1}{3} \
{\rho}^2 \))\,,
\label{Hbub}
\end{equation} 
where both $\lambda_b$ and $\rho =r/a_0$ are expressed in atomic units. 
 The perturbed atomic state is now a mixture of S and D states, with no component of the orbital
angular momentum along the $\vec n$ axis. The spin Hamiltonian is modified and an anisotropic hyperfine
interaction is induced by the D state admixture. The new spin Hamiltonian reads:
\begin{equation}
{\widetilde H_{spin}}= A_{\perp}\;\vec s \cdot
\vec I+ (A_{\parallel}-A_{\perp})(\vec s\cdot \vec n)(\vec I\cdot \vec n) -g_s \mu_B
\vec s \cdot \vec B_0 - \gamma_I \vec I \cdot \vec B_0 \,.
\end{equation}
It is easily verified that, if $ \vec{n} $ lies in  the $(\vec B, \hat u)$ plane, with non-zero 
components along both $\vec B$ and $\hat u$, this perturbed atomic Hamiltonian is no longer invariant under
the transformation  $\Theta$. 
  
Another possible method for breaking the symmetry of $H_{spin}$ will be presented in section 4. 
        
\subsection{Strong magnetic field limit ($\gamma_s B_0
\gg A_{\perp}, A_{\parallel}$)}

The anisotropy axis is defined as:
\begin{equation}
\vec  n = \cos{\psi} \;\hat z\;+ \sin{\psi} \; \hat x \,.
\end{equation}

Let us consider the nuclear spin Hamiltonian associated with the restriction of $H_{spin}$ to the
electronic eigenstate ${\cal E}(\tilde{ n_s}, m_s) $ perturbed by the quadrupolar potential
$H_b(\vec n)$: 
\begin{eqnarray}
H^{eff}_{(m_s)} &= 
& A_{\perp} m_s I_z+m_s (A_{\parallel}-A_{\perp}) (\sin{\psi} \cos{\psi} 
I_x + \cos^2{\psi} I_z) +\gamma_s B_0  m_s - \gamma_I B_0 I_z  \nonumber \\ 
 &=& m_s \lbrack
\gamma_s B_0  + I_z (A_{\perp} \sin^2{\psi}+A_{\parallel}
\cos^2{\psi} -\frac{\gamma_I B_0}{m_s}) + I_x (A_{\parallel}-A_{\perp}) \sin{\psi}
\cos{\psi}\rbrack \nonumber
\end{eqnarray}
 $H^{eff}_{(m_s)}$ is identical to the Hamiltonian seen by an isolated nucleus coupled to an effective
magnetic field, $\vec B^{eff}(m_s)$, having the following components:
\begin{eqnarray}
 B_x^{eff} &=&  -(A_{\parallel}-A_{\perp}) \sin{\psi} \cos{\psi} \;\frac{m_s}{\gamma_I} 
\nonumber\\
 B_y^{eff} &=& 0  \nonumber \\
 B_z^{eff} &=& B_0 - (A_{\perp}\sin^2{\psi} + A_{\parallel} \cos^2{\psi}) \;
\frac{m_s}{\gamma_I}\,, 
\end{eqnarray}
 or equivalently :
$$  B_x^{eff}= B^{eff} \sin{\alpha}, \;\;\; B_y^{eff} = 0, \;\;\; B_z^{eff} = B^{eff} \cos{\alpha}, $$
where
$$ \tan{\alpha}= \frac{ (A_{\parallel}-A_{\perp}) \sin{\psi} \cos{\psi} \;m_s}{-\gamma_I B_0 +
(A_{\perp} \sin^2{\psi} + A_{\parallel} \cos{^2{\psi})\; m_s}} \;. 
$$
In other words, the direction $\hat z^{eff}$ of $\vec B^{eff} $ can be  deduced from the $\hat z$ axis
by a rotation ${\cal R}(\hat y,\alpha)$ by an angle $\alpha$ around the $ \hat y$ axis. Hence, the eigenstates of
$H^{eff}_{(m_s)}$ are
$\vert m_s  \; \tilde {m_I} >$, where $\tilde{m_I} $  now stands for the {\it z}-component of the
spin $\widetilde{\vec I~}$ resulting from $\vec I$ through the rotation ${\cal R}(\hat y,-\alpha)$. 
$$\widetilde I_z = \widetilde{\vec I~} \cdot \hat z = \vec I \cdot {\cal R}(\hat y,\alpha) \hat z= \cos{\alpha}\;
I_z + \sin{\alpha } \; I_x  \,.$$ 
We can now compute the linear Stark shift associated with the Hamiltonian $H_{st}^{pv}$ given by Eq.(\ref{HStark}),
supposing the $\vec E$ field directed along the $\hat y$ axis:
\begin{eqnarray}
 \Delta E_{st} &=& \langle m_s \; \tilde{m_I} \vert d_I E \; s_z I_x  \vert m_s \; \tilde{m_I} \rangle
\nonumber
\\  &=& d_I E\; m_s \langle \tilde{m_I} \vert I_x \vert \tilde{m_I} \rangle  \nonumber \\
&=& d_I E\; m_s \langle  m_I \vert I \cdot {\cal R}(\hat y,\alpha) \hat x \vert m_I \rangle \nonumber \\
&=& - d_I E\; m_s \; m_I \sin{\alpha} \,. 
\end{eqnarray}

 If we suppose  $\gamma_I B_0  \ll A_{\parallel},\;A_{\perp} \ll \vert \gamma_s \vert B_0   $ and
$\vert A_{\parallel}-A_{\perp} \vert  \ll A_{\parallel}+A_{\perp} $, we obtain:
$$ \tan{\alpha} \approx \frac{A_{\parallel}-A_{\perp}}{\frac{1}{2}(A_{\parallel}+A_{\perp})} \sin{
\psi} \cos{\psi}\;\approx \sin {\alpha} \,, $$ which yields the
simplified expression:
\begin{equation}
 \Delta E_{st} = - d_I E \; m_s m_I \frac{A_{\parallel}-A_{\perp}}{A_{\parallel}+A_{\perp}} \sin{2
\psi} \; .
\end{equation}
 In this approximation, $\Delta E_{st} $ can be considered as a modification of
the hyperfine constant linear in the applied electric field.

In order to show up the transformation properties of $\Delta E_{st}$, it is useful to express this
last result in terms of the two fields, $\vec
E$ and $\vec B$, and the unit vector $\hat n$ which defines the anisotropy axis:
\begin{equation}
\Delta E_{st}=- 2 d_I m_s m_I \frac{(\hat n \cdot \vec B_0 )(\hat n \cdot \vec E \wedge \vec
B_0)}{\vec B_0^2}
\; \frac{A_{\parallel}-A_{\perp}}{A_{\parallel}+A_{\perp}} \,.
\label{Delst}  
\end{equation}
 From this expression, it is clearly apparent that the linear shift breaks space reflexion symmetry but
preserves time reversal invariance. It differs from the P and T violating  linear Stark
shift arising from an electron EDM by the fact that it cancels out  when the quadrupolar
anisotropy of the ground state vanishes. 
It is also obvious from Eq. (\ref{Delst}) that, in the strong field limit, the size of the Stark shift depends 
only on the orientation of $\vec B$ relative to $\vec E$ and $\vec n$ and not on the strength 
of the magnetic field. Figure 2 represents two mirror-image configurations of the experiment.   
\begin{figure}
\centerline{\epsfxsize=100mm \epsfbox{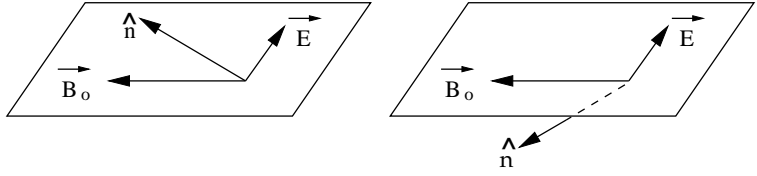}}
\caption{ \footnotesize Two mirror-image  and T-reversal symmetric experimental configurations
corresponding to opposite values of the pseudoscalar 
$ (\hat n \cdot \vec B_0 )(\hat n \cdot \vec E \wedge \vec
B_0)/\vec B_0^2 $.    }
\end{figure}  
 
 \subsection{Limit of low magnetic fields and small anisotropy} 
 We now consider the limit $ \vert A_{\parallel}-A_{\perp} \vert \ll
\gamma_s B_0 \ll A_{\perp}, A_{\parallel} .$

The linear Stark shift can be computed by using second order perturbation theory. $H_{spin}$
is perturbed by both $H_{pv}^{st}$ and $H_b( \vec{n} )$, the latter being responsible for the anisotropy
contribution to $ H_{spin}$,  i.e. $ ( A_{\parallel}-A_{\perp})
(\vec s\cdot \hat n) (\vec I\cdot  \hat n) $. The fields $\vec B_0 $ and $\vec E$ are still taken parallel 
to $\hat  z $ and $\hat y$ respectively. We find:
$$\Delta E_{st}(F, M) = 2 (A_{\parallel}-A_{\perp}) d_I E \cos{\psi} \sin{\psi} \times$$
$$\noindent \times \sum_{F'\not=F, M' }{ \frac{ \langle F \; M \vert s_z I_x + s_x I_z\vert F' \; M'  \rangle  
\langle  F' \; M' \vert s_z I_x - s_x I_z \vert F\; M  \rangle}{ E_{FM} - E_{F'M'}}} \,.$$ 

Since the operator $\vec s\wedge \vec I$ is identical to the commutator $[\vec F^2, -\frac{i }{2}
\vec s]$, we see that only the hyperfine states $F'\not= F$ with $M'=M \pm 1$ contribute to the
 sum. Therefore, in the energy denominator we can neglect the Zeeman contribution which is small
compared to the hyperfine splitting and, in the sum, we can factorize out the energy denominator 
$2(F-I) A_{\parallel}(I+ \frac{1}{2})
$. Since $F^\prime = F $ does not contribute, the resulting sum can be performed using a closure relation:  
 \begin{equation}
\Delta E_{st}(F, M)=\frac{(A_{\parallel}-A_{\perp}) d_I E}{2(F-I) A_{\parallel}(I+ \frac{1}{2})}  \sin{2 \psi} 
\langle F \; M
\vert (s_z I_x + s_x I_z) 
 (s_z I_x - s_x I_z) \vert F\; M \rangle\,.  
\end{equation}
\begin{figure}
\centerline{ \epsfxsize=130mm \epsfbox{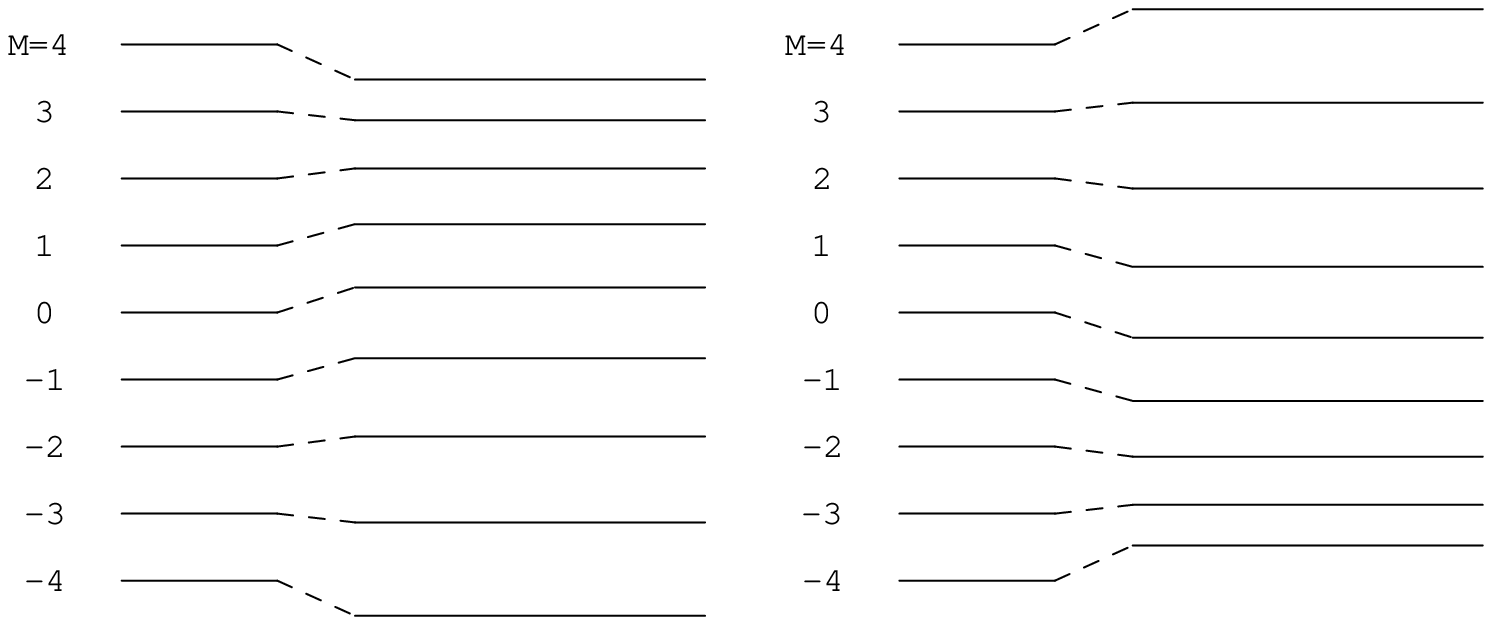}} 
\vspace{10mm}
\centerline{\hspace{0mm} \epsfxsize=130mm \epsfbox{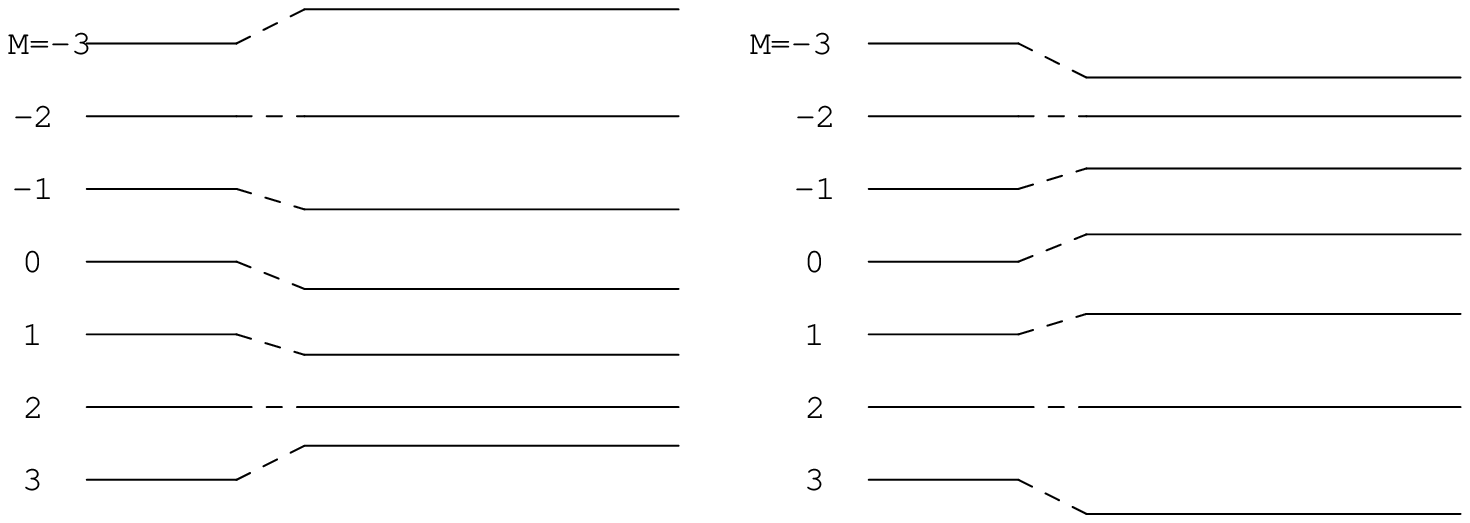}} 
\caption{ \footnotesize Energies of the different F, M states of the ground configuration of $^{133}$Cs showing
the linear Stark shift largely magnified. The two figures correspond to situations realizing opposite signs of the
pseudoscalar
${\cal P}$ defined in the text (Left: ${\cal P}> 0 $; right: ${\cal P}< 0 $). Top: F=4; bottom: F=3.   }
\end{figure}  
Using standard properties of  spin $ 1/2$ matrices, we can transform the diagonal
matrix element above into $\frac{1}{4} \langle  F\;M\vert (I_x^2 - I_z^2) \vert F\; M\rangle $.  
Once taken into account
the axial symmetry of the unperturbed atomic state, it still
simplifies to
$\frac{1}{8} \langle F\;M\vert { (\vec I}^{\,2} - 3 I_z^{\, 2}) \vert F\;M \rangle $.

We arrive at the final expression:
 \begin{equation}
\Delta E_{st}(F, M)= k(F, M)\; \frac{A_{\parallel}-A_{\perp}}{A_{\parallel}+ A_{\perp}} d_I E 
\sin{2 \psi} \;,
\label{delE}
\end{equation}
where 
\begin{equation}
k(F,M)=  2 (F-I) \langle F\;M\vert  \frac{1}{2} ({\vec I}^{\,2} - 3 I_z^{\, 2}) \vert F\;M \rangle / (2I + 1) \,. 
\end{equation}
The Stark shift coefficients $k(F, M) $ for $ ^{133}_{~55}$Cs (I=7/2) are listed in Table~1. 
 
We note that $\Delta E_{st}$ depends on $M^2$ so the linear Stark shifts of the Zeeman  
splittings
$E(F, M)-E(F, M-1)$ have opposite signs for $M > 0$ and $M < 0$ (see figure 3):
$$ E(4, M)-E(4, M-1)  = \hbar \omega_s M / (2 I + 1) + \Delta E_{st}(4, \vert M \vert ) - \Delta
E_{st}(4, \vert M-1 \vert)  \; . 				$$

\noindent As expected, once again the  pseudoscalar
\begin{equation}
 {\cal P}=\frac{A_{\parallel}-A_{\perp}}{A_{\parallel}+A_{\perp}}
\frac{(\hat n
\cdot \vec B_0 )(\hat n \cdot \vec E \wedge \vec B_0)}{B_0^2} \; ,
\end{equation}
 plays an essential role. If
${\cal P}> 0 $, there is a contraction of the Zeeman splittings belonging to the F=4 hyperfine state for 
positive values of
$M$ and a dilatation for negative ones, as shown by Figure 3. The situation is reversed when the sign of
${\cal P}$ is changed.  In the F=3 hyperfine state, splitting
contraction also occurs for $M > 0$ with ${\cal P} > 0$ and for $M < 0$ with ${\cal P} < 0$.  This 
behavior could help to discriminate the linear Stark shift induced by the nuclear helimagnetism from spurious
effects. The largest shift between two contiguous sublevels is expected to occur for the couple  of states $F=3,
M=\vert 3 \vert
\rightarrow F=3, M=\vert 2 \vert$.     

From Table~1 and Eqs. (\ref{dI}) and (\ref{delE}) we predict:
\begin{equation}
\Delta E_{st}(3, 3)-\Delta E_{st}(3, 2) = \frac{75}{64}
\;  \frac{A_{\parallel}-A_{\perp}}{A_{\parallel} + A_{\perp}} \; \sin{2 \psi} \; d_I \cdot E \; , 
\label{Starkshift} 
\end{equation}
with $ \frac{75}{64} d_I \simeq  2.76 \times 10^{-13} \vert e \vert a_0\;.$

As in the strong field limit, we note that the size of $\Delta E_{st}$ depends only on 
the direction of $\vec B$.  
 \begin{table} 
 \begin{center}
\hspace{1 cm}
\begin{tabular}[h]{|c|c|c|c|c|c|}
\hline
 $\vert M \vert $ & 0 & 1 & 2 & 3 & 4 \\   
\hline
 k(4, $\vert M \vert $)  & 15/16 & 51/64  & 3/8   & -21/64  & -21/16 \\   
\hline
 k(3, $\vert M \vert $)  &  -15/16 & - 45/64 & 0 & 75/64 & ~ \\ 
\hline
\end{tabular}
\end{center}
\caption{\footnotesize Linear Stark shift coefficients k(4, M) and k(3, M) of the different F, M substates 
of the natural cesium ground state}
      \end{table}
\subsection{Analogy between this shift and  
 the PV energy shift searched for in
enantiomer molecules}
We would like to stress that from the point of view of symmetry considerations there exists a close
analogy between the linear Stark shift induced by the anapole moment and the energy shift which is
searched for in enantiomer molecules \cite{dau99}. Indeed in the present configuration the three vectors $\vec E,
\vec B$ and $\vec n$ which are non-coplanar are sufficient to place the atom in a chiral environment 
similar to that experienced by an atomic nucleus inside a chiral molecule. Between 
two mirror-image
environments an energy difference is predicted exactly like between two mirror-image molecules.

\section{ Experimental considerations and order of magnitude estimate}
 We now consider an experimental situation which looks like a possible
candidate for the observation of  the linear Stark shift discussed in the previous sections. It has
been demonstrated experimentally
\cite{wei95} that cesium atoms can be trapped in solid matrices of $^4$He. At low pressures, solid
helium cristallizes in an isotropic body-centered cubic (bcc) phase, but also in a uniaxial hexagonal
close packed (hcp) phase. Optically detected magnetic resonance has proved to be a
sensitive tool for investigating the symmetry of the trapping sites. The group of A. Weis
has reported the observation in the hexagonal phase of zero-field magnetic resonance
spectra and magnetic dipole-forbidden transitions which they interpret in terms of a
quadrupolar distorsion of the atomic bubbles \cite{wei98}. 
Particularly relevant here is their observation of the matrix-induced lifting of the Zeeman
degeneracies in zero field. This is attributed to the combined effect of two interactions, the
quadrupolar interaction of the form
$ H_b( \vec{n} )={\lambda}_b \((  (\vec{\rho}\cdot \vec{n})^2  - \frac{1}{3} \
{\rho}^2 \)) $ between the cesium
atom and the He matrix on the one hand, and the hyperfine interaction in the Cs
atom on the other.

Provided that $\vec F^2$ is still 
a good quantum number, it is  easily shown from  general symmetry considerations, 
that the anisotropy of the hyperfine interaction  induced
by the  (hcp)  crystal potential   can be represented,  within a given  hyperfine multiplet,
 by the effective perturbation:  
$$ H_{eff}= C_{eff}(F) \cdot ((\vec F \cdot
\hat n)^2 - \frac{1}{3} \vec F^2) \,.$$
The constants $C_{eff}(F) $ can  be easily  related to the anisotropic hyperfine constants  
appearing in the  spin hamiltonian  ${\widetilde H_{spin}} $ 
introduced in Eq.(\ref{Hbub}) of the previous section:
 $$C_{eff}(F = I \pm 1/2 )= \pm  (A_{\parallel} - A_{\perp})/8\;.$$ 
In a uniaxial crystal, when the atoms are optically polarized along the crystal axis in the 
absence of external
magnetic fields, the lifting of the degeneracy between Zeeman sublevels induced by $
H_{eff}$ should make it possible to drive magnetic resonance transitions between these
levels. One would expect to deduce the
hyperfine anisotropy from the observed spectra. At first sight, the zero-field magnetic resonance spectra
observed by Weis {\it et al.} would seem to match this prediction.
 However, their experiment has been performed in a polycristalline (hcp) sample.
The effects observed in this situation result from averaging over the distribution of the 
microcrystal axes. For each microcrystal, there exists a quantization axis, $ \hat{z}$, which diagonalizes
the hyperfine level  density matrix. Immediately a question arises as to   
 the direction  of the quantization axis $ \hat{z} $ with respect to 
the microcrystal symmetry axis  $\vec{n}$.  If the population
differences resulted, say, from  the Boltzmann  factor, then   $ \hat{z}$ would  be along $\vec{n}$,
since in the zero  magnetic field  limit there is no other preferred direction. In such a situation,  there 
would be no difference between the spectra for a polycrystal and a monocrystal. But in  the 
experimental situation considered here, the population differences are induced by an optical pumping
mechanism which provides a second preferred direction: 
the direction of the photon angular momentum along $ \vec{k} $. The microcrystal density matrix is  then expected 
to  keep  some  memory of the direction of $ \vec{k} $. So, two directions 
 $\vec{n}$ and $ \vec{k} $  compete in the determination of the quantization axis $  \hat{z} $. 
To proceed further, we consider the extreme  case where   $  \hat{z} $ is taken along  $\vec{k}$, 
together with an assumed
isotropic distribution of microcrystal axes. 
It is then  easily seen that the lines associated with the hyperfine anisotropy 
$ H_{eff}$  collapse into a single asymmetric line
  when the average   is performed over 
the polycrystal. Clearly, one at least of the two  preceeding 
assumptions  is too  drastic, most likely the isotropy of the $\vec{n}$  distribution. 
It  is indeed likely  that the optical pumping process is more efficient for
microcrystals having a preferred orientation with respect to the photon angular momentum.
 Such a selection mechanism would then lead to an effective anisotropic distribution of $\vec n$,  
and  in this way a spectrum of separated lines can be recovered. 
 From the above qualitative considerations, 
it clearly follows that the final  interpretation of the  the zero-field resonances   requires
 a detailed analysis of the optical pumping process  for Cs atoms  trapped inside
 deformed bubbles of arbitrary orientation. The corresponding  theoretical investigation is
currently underway in A. Weis's group.

Meanwhile, to plan any experiment, we still need to know about  the  physical origin  and the 
magnitude of the ratio
$ \frac{A_{\parallel}-A_{\perp}}{A_{\parallel}+ A_{\perp}}$, which   governs  the
magnitude of the electroweak linear Stark shift. We are going to
present  now the result of an investigation  which has  led us    both to  a physical 
understanding  and  a reasonably accurate   estimate of the sought after parameter.
We have chosen to devote an appendix to a detailed
 description  of   our semi-empirical  approach,  which consists in  
 relating  the hyperfine  anisotropy to another measured physical  quantity.
Here we shall  give a brief summary of our procedure  and present the final  result.    
  
We start from the remark that there really does exist a mechanism  able  to
generate an hyperfine aniso\-tropy  to first  order in  the 
``bubble'' Hamiltonian  $ H_b( \vec{n} )$. The $nD_{3/2}$ state is indeed mixed 
to the $6S_{1/2} $ state under the effect of
$H_b( \vec{n} )$, and we note then  that the hyperfine interaction 
has non-zero off-diagonal matrix
elements between $S_{1/2}$  and $D_{3/2}$ states. 
In fact, it has been shown previously \cite{bou881} that the
$\langle nS_{1/2}\vert H_{hf}\vert n^{\,\prime }D_{3/2}\rangle$ 
matrix elements are not easy to calculate,
because they are dominated by the contribution coming from many-body effects, 
due to  the existence of an
approximate selection rule which suppresses the single particle matrix element.  
However, as we show  in the appendix,  
the variation of the  matrix elements $\langle n^{\prime} S_{1/2} \vert H_{hf} \vert n^{\prime \prime} D_{3/2}
\rangle $ with respect to the binding energies ${ \cal E }_{n^{\prime} S_{1/2}
}$ and  
${ \cal E }_{n^{\prime \prime} D_{3 /2} }$, -expressed in  Rydberg-  can be reasonably well  predicted   in the 
limit  $\vert { \cal E }_{n^{\prime} S_{1/2} } \vert \, ,
 \, \vert{ \cal E }_{n^{\prime \prime} D_{3 /2} }\vert \ll 1 $. In this way, we are left with a
single parameter  which can be deduced from  the empirical knowledge of
 another physical quantity involving  the same  matrix elements. 
We have in mind  the quadrupolar amplitude $E_2^{hf} $ induced 
by the hyperfine interaction which is present in the cesium 
 $ 6S \rightarrow 7S$ transition in the absence of  a
static electric field \cite{bou882}.   
In order to show the relation between the quantities
 $A_{\parallel}-A_{\perp}$ and $E_2^{hf}$, we express them explicitly in terms of 
the matrix elements $M( n^{\prime },n ) $ given by:
\begin{equation}
M( n^{\prime },n ) =\sum_{n^{\prime\prime} }  \frac{\langle n^{\prime }  S_{1/2}   \vert  H_{hf} \vert 
  n^{\prime\prime} D_{3/2}\rangle  
\langle n^{\prime\prime} D_{3/2}\vert { \rho }^2 \vert  n \, S_{1/2}\rangle } 
{ { \cal E }_{n^{\prime} S_{1/2} }-
{ \cal E }_{n^{\prime \prime} D_{3 /2} } }\, .
\end{equation} 
  The basic formula used  in our  numerical evaluation of
$A_{\parallel}-A_{\perp}$  can be cast  in  a very compact form:
\begin{equation} 
A_{\parallel}-A_{\perp} = - \frac{4\,{\lambda}_b}{ \Delta {\cal{E} }} 
\; \frac{ 2 M(6,6) }{M(7,6)+M(6,7) } \; a_3(7,6) \; Ry
\end{equation}
where $\Delta {\cal{E} }$ is the energy of the $ 6S\rightarrow 7S$ transition and $a_3 \propto \, E_2^{hf} /
\mu_B $ is  the empirical quadrupolar amplitude (see Eq. (A.\ref{M1E2hh}) for a precise  definition). A
second empirical input  is used to determine the coupling constant $\lambda_b$: this is the $S-D$ mixing
coefficient which is obtained from the hyperfine frequency shifts measured by Weis {\it et al.} for
Cs atoms trapped either in the (bcc) or the (hcp) phases \cite{wei98} in the low magnetic field limit.   
 The ratio involving the matrix elements $M( n^{\prime },n )$  is evaluated     
 in the appendix, using  the approximation scheme sketched  above.  Its absolute value is found 
 to lie close to unity. 
 Let us   quote now the final result given by our  semi-empirical method\footnote{This method can be
seen as a generalization of  that used in sec.1.2 to evaluate the static dipole starting from
the empirical knowledge of the transition dipole.}  described  in the appendix:
$\vert \frac{A_{\parallel}-A_{\perp}}{A_{\parallel}+A_{\perp}} \vert = 1.07 \times 10^{-3}$. The
uncertainty is believed not to exceed $20\%$.

For observing the electroweak linear Stark shift discussed in the present paper, it is important to
work with a uniaxial hexagonal crystal. Indeed, in a polycristalline phase, where the individual
crystals are oriented totally at random, the average value of the pseudoscalar ${\cal P}$ taken
over the isotropic distribution of
$\hat n$  is expected to be suppressed  and thus is the Stark shift computed in the previous
section. Although trapping of cesium atoms has not yet been achieved in a monocrystalline hexagonal
phase, the prospect does not look unfeasible
\cite{bal} and a  determination of the magnitude of the hyperfine anisotropy appears to be the first
step to be achieved. Using  $\vert\frac{A_{\parallel}-A_{\perp}}{A_{\parallel}+ A_{\perp}}\vert = 
1.07 \times 10^{-3}$ and Eq.(\ref{Starkshift}), we find that the effective P-odd T-even electric dipole
moment of the trapped cesium atoms associated with the nuclear anapole moment reaches
$ 2.96 \times 10^{-16} \vert e \vert a_0$. 
For comparison, it is interesting to note that this is about three times as large as the Cs EDM  
limit (Eq.\ref{dCsEDM}) to be measured on unperturbed Cs atoms for improving our  present knowledge about a
possible  P-odd T-odd EDM of the electron.
 
\section{ Breaking the free atom symmetry by application of a nonresonant radiation field}        
In this last section we want just to mention another possibility for breaking the atomic Hamiltonian
 rotation symmetry
by other means than  static  uniform 
electric and magnetic  fields. We have in mind the application
of a strong nonresonant radiation field which generates 
 an anisotropic  electron gyromagnetic ratio. In the presence
of an external magnetic field
$\vec B$ it has been shown
\cite{lan70} that the effect of the nonresonant radiation field can be described by the introduction of an
effective magnetic field:
$$ \vec B' = \left( g_{\perp} \vec B + (g_{\parallel} - g{_\perp}) \hat n \cdot \vec B \;\hat n \; \right)/
\sqrt{(g_{\parallel}^2 + g_{\perp}^2)}\; ,$$
where $\hat n$ defines the direction of polarization of the radiation field, $g_{\perp}=g_F $
and $ g_{\parallel}= g_F  J_0(\omega_1/ \omega)$, $J_0$ is the zero-order Bessel
function, and $\omega_1 $ is the Rabi angular
frequency associated with the radiation field.
The above formula suggests the existence of a uniaxial symmetry, but it is valid only within an atomic 
hyperfine multiplet.
 It is clear that  the ``dressing'' by a  nonresonant radiation
field   offers new  possibilities for
placing the atoms in a quadrupolar environment. However, 
it is important to bear in mind that at
least two stringent  requirements must be satisfied if one wants to detect 
an electroweak Stark shift in the ground
state. First, the uniaxial  perturbation has to mix 
the two hyperfine substates, otherwise
the matrix element of
$H_{pv}^{st}$ cancels. Second, it is imperative  to avoid 
a broadening of the transition lines for
allowing precise frequency measurements. We are currently
 investigating how to achieve the proper
conditions in a realistic way. 

\section*{Conclusion}
This paper investigates a way to get around the well known no-go theorem: \\
{\it no  linear Stark shift can be observed in a stationary atomic state unless T reversal invariance 
is broken. }

The perturbation of an atom by the nuclear spin-dependent parity-odd potential 
generated by the nuclear anople moment leads 
to a static electric  dipole moment $ d_I \,\vec s \wedge \vec I  $, which clearly is T-even. 
However, if one considers  an  atom placed in arbitrarily oriented 
electric and magnetic  uniform static  fields $ {\vec B}_0 $ and $ {\vec E}_0 $, 
the quantum  average $ {\vec E}_0 \cdot \langle \vec s \wedge \vec I\rangle $
 is found to vanish. 
This can be  understood by noting that  $ \vec s \wedge \vec I \cdot  {\vec E}_0  $   is odd under the 
quantum symmetry transformation $\Theta$  defined as the product of the time
 reflexion  $ T$ by a space  rotation of $ \pi $ about an axis normal to  a plane parallel to
the fields $ {\vec B}_0  , {\vec E}_0$, while the atomic hamiltonian  stays even. Our strategy to obtain
a linear Stark shift is to break  the  $\Theta$  symmetry while keeping   T invariance.

 As a  possible practical  
realization of such  a situation,  we  have studied the case  of ground state Cs atoms
trapped in a uniaxial (hcp) phase of solid $^4$He, which has been recently the subject 
of detailed spectroscopic studies \cite{wei98}. The required breaking of space
 rotation is provided  by the uniaxal  crystal field.  As a result of  the deformation 
of the atomic  spatial wave function the hyperfine interaction  acquires an anisotropic
part,  which  plays an essential role in the determination of the size of the  linear Stark shift.
 We have performed a  numerical  estimation of the hyperfine anisotropy,  believed to be accurate 
to the $ 20\% $ level, using a semi-empirical method. We use as an input 
the  recent experimental measurement  of the $E_2$ amplitude of the 
$ 6S_{1/2}\rightarrow   7S_{1/2}$  transition
induced in cesium by the hyperfine interaction. We arrive in this way 
 at a  numerical evaluation of the  linear Stark shift induced by the  nuclear  anapole
moment: the expected  effect  is found to be about three times the experimental upper  limit  to be set on the T-odd
Stark shift of free Cs atoms for improving the present limit on the electron EDM.
  
Besides the obvious  remark  that the T-even  Stark shift  studied here could be a possible source of
systematic uncertainty  in EDM experiments designed to reach unprecedented
sensitivity\cite{hun91,com94,hin97,for00}, we believe that there are strong physical motivations for measuring the
Stark shift itself.  First,  it would lead to a direct  measurement of the nuclear
  anapole moment  in absence of any contribution   
coming from the  dominant PV potential due to the weak nuclear charge. It would  also provide
an   evidence  for a  truly static
manifestation of   the electroweak interaction, something which is still lacking. 
Second,  this experiment  would rely  on the measurement of frequency shifts  rather than 
transition amplitudes. While transition probabilities are difficult to measure very  accurately, 
high precision measurements of frequency shifts have already been achieved. 

\section*{Acknowledgements}
We thank Ph. Jacquier for continuous interest in the subject of this work and his encouragements.  
We acknowledge many stimulating discussions with A.~Weis and S.~Kanorsky. We are grateful to M.~Plimmer
and J.~Gu\'ena for careful reading of the manuscript.  

This work has been supported by INTAS (96-334).   

\newpage
\begin{appendix}
\setcounter{equation}{0}
\section*{APPENDIX: 
 Semi-empirical calculation of the hyperfine anisotropy of Cs atoms
trapped inside a $^4$He hexagonal matrix}
 In this appendix we present our evaluation of the hyperfine structure anisotropy $\frac{A_{\parallel}
-A_{\perp}}{A_{\parallel}+ A_{\perp}}$ resulting from the matrix induced bubble deformation of
quadrupolar symmetry, a quantity frequently referred to in this paper.
\subsubsection*{1. Two processes induced by hyperfine mixing} 
Our approach is based on the fact that hyperfine mixing plays quite similar roles in two different
processes. The first process concerns the Cs $6S \rightarrow 7S $ quadrupolar transition amplitude
in zero electric field while the second process deals with the parameter $\frac{A_{\parallel}
-A_{\perp}}{A_{\parallel}+ A_{\perp}}$.

  We start by rewriting  the standard mixed $M_1-E_2$  transition operator in atomic units:
 \begin{equation}   
 T_{ M_1+E_2}= (\vec{\epsilon }\wedge \vec{k} )\cdot \frac{\vec{\cal{M}}}{\mu_B}
-i \frac {1}{2} \Delta {\cal{E} } ( \vec{\rho} \cdot \vec{\epsilon }  )(  \vec{\rho} \cdot  \vec{k} ) \,,
\label{M1E2}
 \end{equation}  
 where $\vec{\rho} $ is the electron coordinate in Bohr radius unit and $  \Delta {\cal{E} } $ is the transition
energy  expressed in Rydberg unit\footnote{The phase difference, $ \pi/2$, between the two
amplitudes expresses the fact that the   magnetic moment
$\vec{\cal{M}}$ and the quadrupole operator behave differently under time reflexion: the first is odd, while the
second is even.}. We are  first   going to study  the perturbation effect  on $ T_{ M_1+E_2} $  caused by the
hyperfine interaction $ H_{hf} $. This phenomenon has been observed experimentally in the  forbidden $
6S_{1/2}
\rightarrow 7S_{1/2}$ transition. It provides a useful calibration amplitude in cesium parity
violation  experiments. To analyse the experimental results, it was found convenient, to
introduce the effective transition  operator $T_{hf} $ acting upon the tensor products of  the electron spin and
nuclear spin states:
\begin{equation}
T_{hf}= i a_2( n^{\prime},n ) \, ( \vec{s}\wedge \vec{I})\cdot  (\vec{\epsilon }\wedge \vec{k} )+
i a_3(  n^{\prime},n) \, ( (\vec{s} \cdot \vec{k}) (\vec{I}\cdot\vec{\epsilon } )+(\vec{s}
\cdot\vec{\epsilon } ) (\vec{I}\cdot\ \vec{k})) \,.
\label{M1E2hh}
\end{equation}

  The second  physical process   to be analysed in this section  is not  at first sight  closely connected but
happens to be  described by the same  formalism. This will allow us to establish a very useful
connection between   measurements coming  from rather different experimental situations. Recently optical
pumping   has been observed  with cesium atoms trapped inside an hexagonal matrix of solid helium \cite{wei98}.
Among  the new effects to be expected, we have seen earlier in this paper  that the existence of an anisotropic
hyperfine structure opens the possiblity of observing a linear Stark shift induced by the nuclear anapole
moment, an effect which cannot exist for an atom in a spherically symmetric environment. It is  
known  that in
the bubble  enclosing the cesium atom there is a  small overlap  between the cesium and the
helium orbitals \cite{kan98}. As a consequence, the axially symmetric crystal  potential inside  the bubble can be
 well approximated by a regular solution of the Laplace equation:
\begin{equation}  
H_b( \vec{n} )={\lambda}_b (\frac{e^2}{ 2 a_0} )\((  (\vec{\rho}\cdot \vec{n})^2-\frac{1}{3} \
{\rho}^2 \)) \,.
\label{hbub}
\end{equation} 
The perturbation of the hyperfine interaction by the bubble quadrupole potential $H_b( \vec{n} ) $
induces an anisotropic hyperfine structure  for  cesium $ nS_{1/2} $ states. This is described
 by  the effective Hamiltonian:
\begin{equation} 
H_{hf}^{anis}= (A_{\parallel}-A_{\perp} ) \(( (\vec{s}\cdot \vec{n} )(\vec{I}\cdot\vec{n}) -
\frac{1}{3} \vec{s}\cdot\vec{I} \))\,.
\label{hfanis}
\end{equation}

We  present now the basic  formulas which allow the computation of the parameters  relevant for the two 
physical problems in hand. They will be given in such a way as to exhibit their close
anology. We have chosen to use the Dirac  equation  formalism. Besides the fact that formulas are more
compact, it is well known that relativistic corrections play an important role in cesium  hyperfine structure
computation.
 Neglecting the contribution of the quadrupole nuclear moment of the Cs nucleus\footnote{As shown in ref
\cite{bou882}, the quadrupole contribution for $^{133}$Cs plays a negligible role in the effects discussed in this
appendix.}, the hyperfine hamiltonian is written as:
 \begin{equation}
H_{hf}= \vec{I}\cdot\vec{\cal{A}}\, , 
\end{equation}

\begin{equation}
\vec{\cal{A}}=C_{hf}\frac{\vec{\alpha} \wedge \vec{\rho}  }{ {\rho}^3} +
\delta  {\vec{\cal{A}} }^{(1)}(\vec{\rho},{\vec{\rho}^{\;\prime}} ) + . . .\; .
\end{equation}
 The first term gives the hyperfine interaction  of the valence electron treated as a Dirac particle;  the
second  represents the non-local modification of the hyperfine interaction induced by the excitation
of  core electron-hole  pairs to lowest order and the  dots stand for higher order contributions\footnote{ An
explicit construction  of $\delta  {\vec{\cal{A}} }^{(1)}(\vec{\rho},{\vec{\rho}^{\;\prime}} )$
together with a resummation of an  infinite set of higher order terms, 
within  the many body field theory formalism, is given  in reference \cite{bou81}. See also \cite{fla89} for more
advanced analysis.}. It has been shown in reference
\cite{bou881} that  the off-diagonal  matrix  element  
$ \langle   n^{ \prime \prime}  D_{3/2} \vert \frac{\vec{\alpha}\wedge \vec{\rho}  }{ {\rho}^3} \vert n S_{1/2}
\rangle $
 is strongly suppressed by an approximate selection rule which does not apply to the many-body non 
local operator $\delta  {\vec{\cal{A}} }^{(1)}(\vec{\rho},{\vec{\rho}^{\;\prime}} )$. An evaluation of the 
 latter contribution led to a semi-quantative  agreement with the experimental measurements of
 the ratio  $ a_3(7,6)/a_2( 7,6)$, while the single particle result is too small by about two orders
of magnitude.



To obtain an estimate of the ratio $ (A_{\parallel}-A_{\perp} ) / a_3( 7,6)$ it  is convenient 
to introduce  the cartesian  tensor operator  $ T_{i_1i_2i_3}(E)$.  This object  appears naturally in 
the lowest-order pertubation expressions for the quantities of interest: 
\begin{equation}
T_{i_1i_2i_3}(E) = {\cal A}_{i_1} \,  G_{3/2}^{+}(E)\,( {\rho}_{i_2}  {\rho}_{i_3}-\frac{1}{3}\,
\delta_{i_2,i_3}\, {\rho}^2) \,,
 \end{equation}
where the indices  $  i_1 \, ,\, i_2 \, ,\, i_3 $ take any value between 1 and 3.  The scalar operator 
$  G_{3/2}^{+}(E)$  is the  atomic Green function  operator restricted to  the subspace of $ D_{3/2}$ 
configurations (total atomic angular momentum J=3/2 and positive parity).
  We now proceed to isolate in  $T_{i_1i_2i_3}(E)$  the part  transforming  as a  vector;
 this is the only part to survive after the operator is sandwiched between the projectors  $P(n^{\prime} S_{1/2})
$  and   $ P(n S_{1/2} ) $. This operation is achieved by a decomposition of  $T_{i_1i_2i_3}(E)$ 
into a traceless  tensor $\bar{T}_{i_1i_2i_3}(E)$ and a remainder \cite{group}:
\begin{eqnarray}
T_{i_1i_2i_3}(E) &= &\bar{T}_{i_1i_2i_3}(E)+\frac{3}{10}\,\(( {\delta}_{i_1,i_2}\, T_{\alpha \alpha
i_3}(E)  +
 {\delta}_{i_1,i_3}\, T_{\alpha \alpha i_2 } (E) \))  \nonumber \\
&  &-\frac{2}{10} \, {\delta}_{i_2,i_3}\,T_{\alpha \alpha i_1}(E) \,, 
\label{irreduc}
\end{eqnarray}
where we have used the fact that 
$T_{\alpha \alpha i } = T_{\alpha i \alpha } $ and $T_{i \alpha \alpha }= 0 $.
 It is a simple matter to verify from the above equation  that  we have indeed
 $\bar{T}_{\alpha \alpha i_3} =\bar{T}_{\alpha i_2\alpha } =\bar{T}_{i_1\alpha \alpha } =0$. 
The fully symmetric part of the traceless tensor $\bar{T}_{i_1i_2i_3}^S (E)$ is easily identified with 
an octupole spherical tensor having seven independant components. By a simple counting argument, the
left over term  is seen to have  five  components; it is to be identified with the quadrupole tensor which
appears in the  full decomposition of $ T_{i_1i_2i_3}(E) $ into irreducible representations of the
rotation group $ O(3)$. Let us have a look at the vector  operator, $ \vec{V} $,  the components   of which
appear  in the right hand side of Eq.(A.\ref{irreduc}): 
$$ 
\vec{V}= (\vec{\cal{A}} G_{3/2}^{+}(E)\cdot  \vec{\rho}) \, \vec{\rho}-
\frac{1}{3} \vec{\cal{A}}  G_{3/2}^{+}(E) {\rho}^2 \,.   
$$
The  second term in the above expression  does not contribute when it  acts upon an $ n S_{1/2} $ state
so, we are led for our purpose  to introduce the vector operator  $ \vec{{\cal T}}( n^{\prime},n)   $
\begin{eqnarray}
\vec{{\cal T}}( n^{\prime},n) &= &\frac{3}{10}  P(n^{\prime} S_{1/2} ) \(( \vec{\cal{A}} 
G_{3/2}^{+}(E_i ) \cdot  \vec{\rho}) \, \vec{\rho}
 +( h.c , E_f \rightarrow E_i )\))  P(n S_{1/2}) \nonumber \\
    & =& \gamma(n^{\prime},n)  \,\vec{s} \,, 
\label{gamma}
\end{eqnarray}
   where $ E_f$ and $ E_i$ are respectively the binding energies of the  
$ n^{\prime} S_{1/2}$  and $ n S_{1/2} $  atomic states. The  second line of the above equation
follows  directly from the Wigner-Eckart theorem applied to a vector operator.

In order  to calculate $ a_3 (n^{\prime},n) $ we have to perform the contraction
 of  $ I_{i_1} {\epsilon}_{i_2} k_{i_3} $  with   the tensor:
$$ F_{i_1i_2i_3}= P(n^{\prime} S_{1/2} ) \(( T_{i_1i_2i_3}(E_f)+( h.c. , E_f \rightarrow E_i )  \)) P(n S_{1/2}) \,.$$
Using Eq.(A\ref{irreduc})  and (A\ref{gamma}) $,  F_{i_1i_2i_3} $ can be cast  into the simple   form :
 $$
F_{i_1i_2i_3}=
\gamma(n^{\prime},n)  \(( {\delta}_{i_1,i_2}\, s_{i_3}+ {\delta}_{i_1,i_3}\, s_{i_2}-\frac{2} {3}{\delta}_{i_2,i_3}\,
s_{i_1} \)) \,.
$$ 
The  required index  contraction with the tensor $I_{i_1} \epsilon_{i_2} k_{i_3}$ is  now easily performed  and one
obtains directly  $ a_3(n^{\prime} ,n)$, up to a prefactor whose value is found by identification
with Eq.(A.\ref{M1E2}):
\begin{equation}
a_3(n^{\prime} ,n)= -\frac {1}{2} \Delta {\cal{E} }\,\gamma(n^{\prime},n)\,. 
\label{E2hf}
 \end{equation}   
 To calculate the hyperfine anisotropy $A_{\parallel}-A_{\perp}$, we follow the same lines  but this time the
contraction involves  the tensor $  I_{i_1} \(( {n}_{i_2} n_{i_3} -\frac{1}{3} {\delta}_{i_2,i_3} \))$, the prefactor is
fixed by comparison with Eq.(A.\ref{hbub})  
 and the exchange $i_2 \leftrightarrow  i_3 $ leads to two identical contributions. Hence,    
\begin{eqnarray} 
A_{\parallel}-A_{\perp} &=  & 2 \,{\lambda}_b (\frac{e^2}{ 2 a_0} ) \,\gamma(n,n)  \\
          &=& - (\frac{e^2}{ 2 a_0} ) \, \frac{4\,{\lambda}_b}{ \Delta {\cal{E} }} 
\, \frac{\gamma(n,n)}{\gamma(n^{\prime},n)}\, a_3 (n^{\prime},n) \,. 
\label{anisA}
\end{eqnarray}
The expression (A.\ref{anisA}) looks to us a good starting point for numerical evaluation of 
$A_{\parallel}-A_{\perp}
$: besides the fact that several sources of uncertainties in the evaluation of  ${\gamma(n^{\prime},n)}$ are
eliminated in the ratio $\frac{\gamma(n,n)}{\gamma(n^{\prime},n)}$, it lends itself to the use of
empirical information. One may note, here, a certain similarity with Eq. (7) of sec 1.2 used for the
evaluation of the permanent dipole, $d_I$. 

\subsubsection*{2. Numerical evaluation} 
 We  proceed now to a    numerical evaluation of  $A_{\parallel}-A_{\perp} $ 
  in three steps,  starting   from formula (A.\ref{anisA}). 

The  numerical value of the $ 6S_{1/2}\rightarrow   7S_{1/2}$ quadrupole amplitude $a_3 (7,6)  $ 
 is readily obtained from   measurements \cite{bou882,ben99} of  the ratio 
$$a_3 (7,6)/a_2 (7,6)= E_2/M_1^{hf} =  ( 5.3 \pm 0.3) \times 10^{-2}\, ,$$  
 combined with a  precise theoretical evalution of the magnetic dipole amplitude\footnote{
The theoretical   method used to get $ M_1^{hf}$ is based upon the factorization rule:
$\langle 6S \vert  H_{hf}\vert 7S\rangle = \sqrt {\langle 6S \vert  H_{hf}\vert 6S\rangle  \langle 7S \vert  H_{hf}\vert
7S\rangle } $.
 This rule  was first established
with an accuracy of a few parts in $10^3$  in ref.\cite{bou881}. 
It has been confirmed  by a direct  many-body relativistic computation 
\cite{joh99} of  $\langle 6S \vert  H_{hf}\vert 7S\rangle $, accurate  to the $1\% $ level. 
More recently the  validity of the  rule 
  has  been  pushed to the level of a fraction of $10^{-3}$ \cite{fla00}.   }
$ a_2 (7,6) = -\frac{ M_1^{hf}}{2\mu_B} = -0.4047  \pm  4 \times 10^{-4}$.  We obtain finally:
\begin{equation}
a_3 (7,6)= ( 2.14 \pm 0.12)\times 10^{-7}\, .
\end{equation}

    The second step is  the numerical estimate of the ratio $ \gamma(6,6) / \gamma(7,6) $. This is more delicate
and requires an assumption which has been shown to work in  similar situations. To begin with, we have
addressed the question\footnote{Arguments similar to those given below  and in
references \cite{bou881,bou86} are developed in \cite{fla00}.} of the origin and size of the variations of the
off-diagonal matrix elements $
\langle  nS_{1/2}\vert H_{hf}\vert n^{\prime \prime} D_{3/2}\rangle $ with the radial quantum numbers $ n$ and $
n^{\prime \prime}$. It is of interest to remind  that
a  very precise answer to  this  question has been already obtained in the case of cesium single particle
matrix  elements $ \langle  nL_{J}\vert H_{hf}^{sp}\vert n^{\prime } L_{J}\rangle  $ with $ L= 0  \;{\rm or}\;1 $ and
with $n \;{\rm and} \; n^{\prime }$ referring to the radial quantum numbers of any pair of valence states. For
simplicity, we are going to express the answer within a non-relativistic formalism, but it should be borne in mind that
all of what is said holds true within  a relativistic framework. It is convenient to introduce  the 
notion of overline matrix  elements such as those
computed with radial wave functions
$\overline{ R_{nlj} }(\rho) $ which have a starting  coefficient at the origin equal to unity instead of a
unit norm\footnote{ The wave function $\overline{ R_{nlj} }(\rho) $ is known to be 
 an analytic function of the energy. This property is the starting   point 
of the quantum defect theory.}. More explicitly we can write: 
\begin{equation}
   \overline {\langle  nL_{J} \vert H_{hf}^{sp} \vert n^{\prime } L_{J}\rangle  } =
\frac{  \langle  nL_{J} \vert H_{hf}^{sp} \vert n^{\prime } L_{J}\rangle   }{  A_{ n l_{j} } \, A_{ n^{\prime } l_{j}
} } \,,
\end{equation} 
 where  $ A_{ n l_{j} }= \lim_{ \rho \to 0} {\rho}^{- l}  R_{nlj} (\rho)  $ is the starting coefficient of the space
normalized wave function. (In the relativistic case the above  condition is  replaced   by  energy independent
boundary conditions  imposed on the Dirac radial wave functions at the nuclear radius). It  was found in
references \cite{bou881,bou86} that the overlined  matrix elements are independent of the valence 
orbital  radial quantum numbers $ n$ and
$n^{\prime }$ to better than $ 10^{-4}$   for $ S_{1/2} $ states  and   better than $ 10^{-3}$   for $ P_{1/2} $ 
states. This result is understood  by noting that, in the domain of the $ \rho $  values relevant
 for the   evalutation of the matrix elements of 
 $ \frac{\vec{\alpha} \wedge \vec{\rho}  }{ {\rho}^3} $  for $ S_{1/2}$  and $  P_{1/2} $
 states, the potential energy is  larger than valence binding energies by
more than three orders of magnitude. 
This implies that, {\it in this domain, the overlined radial
wave functions have no dependence upon  the binding energy or equivalently upon the radial quantum
numbers of the valence orbitals}.

The above  argument has to be reconsidered for the  lowest order
many body correction  involving the matrix element of the non local operator:
$ \delta  {\vec{\cal{A}}}^{(1)}(\vec{\rho},{\vec{\rho}^{\;\prime}} ) $. 
The relevant domain of $ \rho  $ values is now 
determined  by  the  ``radii'' of the  core  outer orbitals involved in the computation, which in the case 
of $ S_{1/2} $ and  $ P_{1/2} $ matrix elements are $ 5s\, , \, 5p \; $,  while in the case of  
the off-diagonal matrix element $ \langle  nS_{1/2} \vert  \delta  {\vec{\cal{A}}}^{(1)}\vert n^{\prime }
D_{3/2}\rangle$  only  $ 5p\, $  is relevant. 
 We  measure  the variation of the overlined matrix elements   
$  \overline{  \langle  nL_{J}\vert H_{hf}^{mb}\vert n^{\prime } L^{ \,\prime}_{J^{\prime}} \rangle } $
with the  valence state binding
energies  $ { \cal E }_{nL_{J} } $ by the parameters $ {\delta }_{L_{J} }$  defined as their logarithmic
 derivative 
with respect to $ { \cal E }_{nL_{J} } $, (here $H_{hf}^{mb}$  stands  for  the many-body modification to the 
hyperfine  interaction). From  results of references \cite{bou881,bou86},  we can infer the relative variation of 
$\overline{
\langle n L_{1/2} \vert  \delta  {\vec{\cal{A}}}^{(1)}\vert n^{\prime } L_{1/2}\rangle } $ for $ L=0,1$ 
and we arrive to  the values 
${\delta }^{(1)}_{S_{1/2} }= -0.12 $ and ${\delta }^{(1)}_{P_{1/2} }= -0.30 $. The fact 
that  $-{\delta }^{(1)}_{P_{1/2} } $ is about three times larger than  $-{\delta }^{(1)}_{S_{1/2} }$
is coming from the fact that  $ P $ state  binding  energies  have   to be compared   with 
the  potential energy minus the centrifugal energy. Let us,   
now, consider the more difficult case of the S-D   
off-diagonal  matrix elements
 $ \overline{   \langle   n^{\prime }S_{1/2} \vert 
 \delta  {\vec{\cal{A}}}^{(1)}\vert n^{\prime\prime} D_{3/2}\rangle   }$.   The corresponding 
parameter  ${\delta }^{(1)}_{S_{1/2} }$ is expected to be somewhat larger
in absolute value, due to the fact that the  relevant $ 5p $ orbital is less tightly bound
 than the $5s$ orbital which gives the dominant contribution to the $  S_{1/2} $ diagonal matrix element. 
The relative variation  versus the
$ D_{3/2} $ energy  is expected to be on the order of few units,  since the  centrigugal barrier is three times
higher than in the case of $ P $ states. This expectation is borne out by  a preliminary  estimate which gives 
 ${\delta }^{(1)}_{D_{3/2} } \sim -3$.

  We proceed now  to a numerical evaluation of the ratio $ r_{anis}=\gamma(6,6)/\gamma(7,6) $,
leaving, for the moment,   ${\delta }_{ S_{1/2}}$ and $ {\delta }_{ D_{3/2}  }$ as free parameters. 
 As an intermediate step, we compute  the quantities  $ M( n^{\prime },n ) $, written as   
sums over the intermediate $ n^{\prime\prime} D_{3/2} $ states :
\begin{equation}
M( n^{\prime },n ) =\sum_{n^{\prime\prime} }  \frac{\langle n^{\prime } S_{1/2}  \vert  H_{hf} \vert 
  n^{\prime\prime} D_{3/2}\rangle  
\langle n^{\prime\prime} D_{3/2}\vert { \rho }^2 \vert  n \; S_{1/2}\rangle } { { \cal E }_{n^{\prime} S_{1/2} }-
{ \cal E }_{n^{\prime \prime} D_{3 /2} } }\,.
\end{equation}
The ratio $ r_{anis}$ is given in terms of $M( n^{\prime },n )$ by the following formula :
\begin{equation}
 r_{anis}=\gamma(6,6)/\gamma(7,6)= \frac{ 2 M(6,6) }{M(7,6)+M(6,7) } \, .
 \end{equation}

An explicit numerical computation of $ r_{anis}$ has been performed according to the following procedure. 
First, any  binding energy 
 independent factor  appearing  in $ M( n^{\prime },n ) $ is dropped since it disappears
in the ratio. This is indicated below by the symbol $\propto $. The sum $ \sum_{n^{\prime\prime} } $   appearing
in  $M( n^{\prime },n )$ is limited to
$ 5 \leq n^{\prime\prime} \leq 8 $. The  set of the quadrupole matrix
element $\langle n^{\prime\prime} D_{3/2}\vert { \rho }^2 \vert  n \; S_{1/2}\rangle $ were 
obtained  by a relativistic version of the Norcross model. In order to test the sensitivity of the result to
quadrupole amplitudes, we have also used a set  calculated by an extension of the Bates-Damgaard  method. The
energy denominators  
 are taken from experiment. The hyperfine matrix  elements $ \langle n^{\prime } S_{1/2} \vert  H_{hf} \vert 
  n^{\prime\prime} D_{3/2}\rangle$, to second order in  the
binding energies are,  given   by  the following formulas:
\begin{eqnarray}
 \langle n^{\prime } S_{1/2}   \vert  H_{hf} \vert 
  n^{\prime\prime} D_{3/2}\rangle &  \propto &  
A_{n^{\prime\prime} D_{3/2} }\, A_{  n^{\prime } S_{1/2}  }  \times  
  \label{hfSDme} \nonumber \\
&  & \(( 1+  {\delta }_{ S_{1/2} } { \cal E }_{n^{\prime} S_{1/2} } + {\delta }_{ D_{3/2}  }
 {\cal E }_{n^{\prime \prime} D_{3/2} } \)) \,,  \\
A_{  n L_{J} }  & \propto &  \,  { (-\cal E }_{n  L_{J} })^{\frac{3}{4} }\,.
\label{startcoef}
\end{eqnarray}
 	In  formula (\ref{hfSDme}), we have dropped, according to the above prescription, the zero
energy limit of the overlined matrix element $ \overline{\langle n^{\prime } S_{1/2}  \vert  H_{hf} \vert 
  n^{\prime\prime} D_{3/2}\rangle}$. Equation  (A.\ref{startcoef}) follows from a result obtained in 
\cite{bou75}, where the Fermi-Segr\' e formula was extended to arbitrary orbital angular momentum states. For
simplicity we have ignored for simplicity a factor  involving the derivative of the quantum defects, which
in the present context would introduce  few percent  corrections. 
 
 We now have  all the elements  needed to calculate the sought after ratio:
\begin{equation}
r_{anis} =\frac{  \gamma(6,6)}{\gamma(7,6)}=-0.8173-0.0255 {\delta }_{ D_{3/2}  }+ 0.1456  {\delta }_{ S_{1/2} }\,.
\end{equation}
    	The negative sign of $ r_{anis} $ can be traced back to the fact that  the $7S_{1/2}$ level 
lies  just in between  $5D_{3/2}$ and $6D_{3/2}$ levels. If we adopt the rough estimate given 
above : a few tens of $ \%  $  for $  -{\delta }_{ S_{1/2} } $ and a few units for $ -{\delta }_{ D_{3/2}  }$,
the first-order energy correction  remains  well below  the $10 \%  $ level,
due in part to a cancellation between the two
correcting  terms. To reduce the absolute value of $ r_{anis} $ by  more
than $ 10 \%$ would require unrealistic values of $-{\delta }_{ D_{3/2}  }$ so we believe  
the estimate of $ r_{anis}= -0.82 \pm 0.10$ to be reasonably safe\footnote{ The validity of the
procedure leading  to this estimate has been checked, to the  $10 \% $ level,   upon a significative subset of
the many-body Feynman diagrams contributing   to  $ \gamma(n^{\prime},n) $.}. 

 The final  step in our evaluation of the hfs anisotropy is devoted to  the empirical determination
of  the coupling constant $ \lambda_b$ appearing in front of the crystal electronic potential. As experimental
input  we are going to use the hyperfine energy shift which is observed for trapped cesium atoms, 
when one passes from the cubic to the hexagonal phase. This shift  is attributed to the effect
of the anisotropic bubble potential $ H_b( \vec{n} )$. We  shall ignore, for the moment, the possible contribution 
of the anisotropic  hyperfine interaction $H_{hf}^{anis}$  and assume that the shift is essentially due
to the renormalization of the $6 S_{1/2} $ component of the atomic wave function by the admixtures 
$ {\alpha}_{n D_{3/2}} $  of the $n D_{3/2}$ states. The corresponding variation of the hyperfine 
splitting  $ \delta W $ is then given by : 
\begin{equation}   
\frac{\delta  W}{ W}=-\sum_{n,J}{ { \vert {\alpha}_{n D_{J} } \vert }^2}=  -\lambda_b^2 {\cal J }_{SD}\,,
\end{equation}
where we have isolated $\lambda_b^2 $ by introducing  the purely atomic quantity ${\cal J }_{SD} $.
 Let us  write down  the explicit expression of ${\cal J }_{SD} $, neglecting spin-orbit coupling and assuming that
$\vec{n}$ lies along the quantization axis: 
\begin{equation}   
{\cal J }_{SD} =   \sum_{n} {\vert   \langle 6 S  
\vert ( \cos ^2\theta-1/3)\rho^2\vert n D \rangle /
( {\cal E}_{6 S}-  {\cal E}_{n D} )   \vert}^2 \,.
\end{equation}   
We limit the sum to $n$ values ranging from 5 to 8. With the same radial quadrupole matrix elements as
before, we obtain the numerical  value: ${\cal J }_{SD}=9512 $. Using the empirical number given in ref \cite{wei98},
$\sqrt{- \delta W/W }=0.035 $, we arrive at the following absolute value of the coupling constant 
$\lambda_b $ (in Ry):   
\begin{equation}   
\vert\lambda_b \vert =    \sqrt{ \frac{-\delta  W}{ {\cal J }_{SD} \, W} }= 0.000359 \,.
\label{lambda}
\end{equation}   
 	It should be pointed out that if  the  crystal axis  $\vec{n}$  is not aligned along the quantization axis, one obtains
   values  of ${\cal J }_{SD}$  smaller than the one quoted above,   so  the value of $ \vert\lambda_b \vert $ should be 
  considered,  strictly speaking, as a lower bound.
At last, we have in hand all the ingredients  needed to perform a numerical evaluation of  
$ \vert A_{\parallel}-A_{\perp} \vert$ from the formula (A.\ref{anisA}) since $ \Delta {\cal E } =0.169$ is taken
directly from experiment: 
  \begin{eqnarray}
\vert A_{\parallel}-A_{\perp} \vert &= &\, \frac{4\,\vert{\lambda}_b\vert}{ \Delta {\cal{E} }} 
\, \frac{\vert\gamma(6,6)\vert}{\vert\gamma(7,6)\vert}\,\vert a_3 (7,6) \vert \; {\rm Rydberg( MHz)  } 
= 4.9  \; {\rm  MHz  }  \\
&=&  1.07 \times  10^{-3} \times ( \vert A_{\parallel}+A_{\perp} \vert ) \,.
\label{finalresult}
\end{eqnarray} 
  As a final topic, we should  discuss the effect of the anisotropic hyperfine interaction  itself on the 
the empirical splitting  $ \delta W $, since this could modify the value of $\vert{\lambda}_b\vert $
 and so play a role in the assessment of the uncertainty
  affecting the result given by Eq. (A.\ref{finalresult}).
Due to this effect, the constant $\lambda_b$ is no longer given by  Eq. (A.\ref{lambda}) 
but rather by a second order equation where
the linear term is associated with the anisotropic hyperfine interaction.  It is convenient to introduce the 
variable $ x= \lambda_b/ \lambda_b^0$ with $ \lambda_b^0 = \sqrt{ -\delta  W /( {\cal J }_{SD} \, W) }$. 
The  equation giving $\lambda_b$  takes  then the simple form: $ x^2-2\,b\,x -1=0 $, where the coefficient $b$
is given by the following formula:
$$
b= \frac{ (A_{\parallel}-A_{\perp} )^{(0)} }{ A_{\parallel}+A_{\perp} } \frac{W}{\delta W}
\frac { \Delta_F \langle  \; s_z I_z  -\frac{1}{3 } \vec{s}\cdot\vec{I} \; \rangle }{ 2\,I+1}\,.
  $$ 
 The superscript $^{(0)}$ indicates  that the  hf  anisotropy is given, up to a well defined 
sign, by Eq. (A.\ref{finalresult}). The symbol $  \Delta_F $ means that one should take the difference 
between the two hyperfine states of the quantum average over which it is applied.
 To obtain an over-estimate  of $b $ we  have assumed that 
optical pumping  works at  its maximum   efficiency so that the microwave transition
takes place between  the 
hyperfine levels (4,4) and (3,3). In this case we obtain $b=0.20$ and 
the two possible solutions  for $\lambda_b$ are :
$$  \lambda_b^{(\pm )}=\pm 3.6 \, 10^{-4} \,(1\pm 0.2) \, .$$
 The actual experimental 
situation  is expected to lie far from the extreme case considered here,  so 
 the difference  between the  two absolute values
 is certainly smaller than the upper limit given by the above calculation.

In conclusion, including all sources of uncertainties, we consider our evaluation 
of  Eq.(A.\ref{finalresult}), $ \vert
\frac{ A_{\parallel}-A_{\perp}}{ A_{\parallel}+A_{\perp} } 
\vert = 1.07 \times 10^{-3} $, as reliable within uncertainty limits of about 20$\%$.
However, if, during hyperfine shift
 measurements $\vec n$ is not aligned along the quantization axis, the central value of 
$\lambda_b $, and therefore that of $\frac{ A_{\parallel}-A_{\perp}}{ A_{\parallel}+A_{\perp} }$, may
be pushed upwards.      
\end{appendix}



\end{document}